\documentclass{paper}
\usepackage{amsfonts,amsmath,latexsym,amscd,graphicx,amssymb}

\newcommand{\be}{\begin{equation}}
\newcommand{\ee}{\end{equation}}
\newcommand{\ba}{\begin{eqnarray}}
\newcommand{\ea}{\end{eqnarray}}

\newcommand{\f}{\frac}

\begin{document}

\title{Particle trajectories beneath small amplitude shallow water waves
in constant vorticity flows }
\author{\normalsize Delia IONESCU-KRUSE\\
\normalsize Institute of Mathematics of
the Romanian Academy,\\
\normalsize P.O. Box 1-764, RO-014700, Bucharest,
 Romania\\
\normalsize E-mail: Delia.Ionescu@imar.ro\\[10pt]}

 \date{}
 
\maketitle

\begin{abstract}
We investigate the particle trajectories in a constant vorticity
shallow water flow over a flat bed as  periodic waves propagate on
the water's free surface. Within the framework of small amplitude
waves, we find the  solutions of the nonlinear differential
equations system which describes the particle motion in the
considered case, and we describe the possible particle
trajectories. Depending on the relation between the initial data
and the constant vorticity,  some particle trajectories are
undulating curves to the right, or to the left, others are loops
with forward drift, or with backward drift, others can follow some
peculiar shapes.

\end{abstract}

\section{Introduction}

The motion of water particles  under regular waves which propagate
on the water's free surface is a very old problem. It was widely
believed that the particle trajectories are closed. After the
linearization of the governing equations for water waves,
analysing the first approximation of the nonlinear ordinary
differential equations system which describes the particle motion,
one obtained that  all water particles trace closed, circular or
elliptic, orbits (see, for example, \cite{debnath},
\cite{johnson-carte}, \cite{kk}, \cite{lamb}, \cite{lighthill},
\cite{sommerfeld}, \cite{stoker} - a conclusion apparently
supported by photographs with long exposure
 \cite{debnath}, \cite{sommerfeld}, \cite{stoker}).

 While in this first  approximation  all particle paths appear to be closed,
 in \cite{cv} it is
shown, using phase-plane considerations for the nonlinear system
describing the particle motion, that in linear periodic gravity
water waves no particles trajectory is actually closed, unless the
free surface is flat. Each particle trajectory involves over a
period a backward/forward movement, and the path is an elliptical
arc with a forward drift; on the flat bed the particle path
degenerates to a backward/forward motion. Similar results hold for
the particle trajectories in deep-water, that is, the trajectories
are not closed existing  a forward drift over a period, which
decreases with greater depth (see \cite{cev}). These conclusions
are in agreement with Stokes' observation \cite{stokes}: "There is
one result of a second approximation which may possible
importance. It appears that the forward motion of the particles is
not altogether compensated by their backward motion; so that, in
addition to their motion of oscillation, the particles have a
progressive motion in the direction of the propagation of the
waves. In the case in which the depth of the fluid is very great,
this progressive motion decreases rapidly as the depth of the
particle considered
increases."\\
 For linearized irrotational shallow water waves, there
are obtained very recently in \cite{io}
 the exact  solutions  of the nonlinear
differential equations system which describes the particle motion.
Beside the phase-plane analysis, the exact solutions allow a
better understanding of the dynamics.   In \cite{io} it is shown
that depending on the strength of underlying uniform current,
beneath the irrotational shallow water waves some particle
trajectories are undulating path to the right or to the left, some
are looping
curves with a drift to the right.\\
 The steady (traveling) linear water waves with constant vorticity were
 studied
recently in \cite{ehrnst}, \cite{ev}. Here linearity means that
the waves are small perturbations of shear flows. The linear
system obtained in this way is solvable. Further,  making a phase
portrait study in steady variables for the nonlinear differential
equations system which describes the particle paths, it is found
that for positive vorticity, the steady wave resembles that of the
irrotational situation, though for large enough vorticity the
particles trace closed orbits within the fluid domain. For
negative vorticity all
the fluid particles display a forward drift.\\
Using the same approach as in \cite{io} within the framework of
linear water waves theory, we investigate in this paper  the
particle trajectories beneath shallow water waves in a constant
vorticity flow. The obtained results and the way in which the
paper is organized are presented below.

Let us now give some references  on the results obtained for the
governing equations without linearization. Analyzing a free
boundary problem for harmonic functions in a planar domain, in
\cite{c2007} it is shown that there are no closed orbits for
Stokes waves of small or large amplitude propagating at the
surface of water over a flat bed; for an extension of the
investigation in \cite{c2007} to deep-water Stokes waves see
\cite{henry}. Within a period each particle experiences a
backward/forward motion with a slight forward drift. In a very
recent preprint \cite{CS},  the results in \cite{c2007} are
recovered by a simpler approach and there are also described all
possible particle trajectories beneath a Stokes wave. The particle
trajectories change considerably according to whether the Stokes
waves enter a still region of water or whether they interact with
a favorable or adverse uniform current. Some particle trajectories
are closed orbits, some are undulating paths and most are looping
orbits that drift either to the right or to the left, depending
on the underlying current. \\
Analyzing a free boundary problem for harmonic functions in an
infinite planar domain, in \cite{CE2} it is shown that under a
solitary wave, each particle is transported in the wave direction
but slower than the wave speed. As the solitary wave propagates,
all particles located ahead of the wave crest are lifted while
those behind have a downward motion.\\
 Notice that there are only a
few explicit solutions to the nonlinear governing equations:
Gerstner's wave (see \cite{gerstner} and the discussion in
\cite{c2001a})), the edge wave solution related to it (see
\cite{c2001b}), and the capillary waves in water of infinite or
finite depth (see \cite{crapper}, \cite{kinn}). These solutions
are peculiar and their special features (a specific vorticity for
Gerstner's wave and its edge wave correspondent, and complete
neglect of gravity in the capillary case) are not deemed relevant
to sea waves.

The present paper is organized as follows.   In  Section 2 we
recall the governing equations for gravity water waves. In Section
3 we present their nondimensionalisation and scaling. It is
natural to start the investigation for shallow water waves by
simplifying the governing equations via linearization. The
linearized problems for an irrotational shallow water flow and for
a constant vorticity shallow water flow are written in Section 4.
We also obtain here the general solutions of these two linear
problems. In the next section we find the  solutions of the
nonlinear differential equations systems which describe the
particle motion in the two cases, and we describe the possible
particle trajectories beneath shallow water waves. We see that
these particle trajectories are not closed. Section 5.1 contains
the irrotational case. Depending on the strength of the underlying
uniform current, the particle trajectories are undulating path to
the right or to left, are looping curves with a drift to the
right, and, if there is no underlying current or the underlying
current is moving in the same direction as the irrotational
shallow water wave with the strength of the current smaller than
2, then, the particle trajectories obtained are not physically
acceptable (Theorem 5.1). In dealing with the linearized problem
not with the full governing equations, we expect to appear
solutions which are not physically acceptable. Section 5.2
contains the case of a constant vorticity flow. Also in this case
the particle trajectories are not closed. Depending on the
relation between the initial data $(x_0,z_0)$ and the constant
vorticity $\omega_0$,  some particle trajectories are undulating
curves to the right, or to the left, others are loops with forward
drift, or with backward drift, others can follow some peculiar
shapes (Theorem 5.2, Theorem 5.3).

\section{The governing
equations for gravity
 water waves}

 We consider a two-dimensional inviscid incompressible fluid in a
constant gravitational field. For gravity water waves these are
physically reasonable assumptions (see \cite{johnson-carte} and
\cite{lighthill}). Thus, the motion of water is given by Euler's
equations
\begin{equation}
\begin{array}{c}
u_t+uu_x+vu_z=-\f1{\rho} p_x\\  v_t+uv_x+vv_z=-\f1{\rho} p_z-g\\
\end{array}
 \label{e}
 \end{equation}
Here $(x,z)$ are the space coordinates, $(u(x,z,t), v(x,z,t))$ is
the velocity field of the water, $p(x,z,t)$ denotes the pressure,
$g$ is the constant gravitational acceleration in the negative $z$
direction and  $\rho$ is the constant density. The assumption of
incompressibility implies the equation of mass conservation \be
u_x+v_z=0 \label{mc}\ee

\noindent  Let $h_0>0$ be the undisturbed depth of the fluid and
let $z=h_0+\eta(x,t)$ represent the free upper surface of the
fluid (see Figure 1). The boundary conditions at the free surface
are constant pressure
\begin{equation}
p=p_0 \,  \textrm{ on } z=h_0+\eta(x,t),
 \label{bc2}
 \end{equation}
 $p_0$ being the constant
atmospheric pressure, and the continuity of fluid velocity and
surface velocity
\begin{equation}
  v=\eta_t+u\eta_x \, \, \textrm{ on }\,
z=h_0+\eta(x,t)
 \label{bc1}
 \end{equation}
On the flat bottom $z=0$, only one condition is required for an
inviscid fluid, that is,
\begin{equation}
 v=0 \, \,
\textrm { on } z=0
 \label{bc1'}
 \end{equation}
Summing up, the exact solution for the water-wave problem is given
by the system (\ref{e})-(\ref{bc1'}). In respect of the
well-posedness for the initial-value problem for
(\ref{e})-(\ref{bc1'}) there has been significant recent progress,
see \cite{shkoller} and the references therein.

A key quantity in fluid dynamics is   the \emph{curl} of the
velocity field, called vorticity. For two-dimensional flows with
the velocity field $(u(x,z,t), v(x,z,t))$, we denote the scalar
vorticity of the flow by \be \omega(x,z)=u_z-v_x
\label{vorticity}\ee Vorticity is adequate for the specification
of a flow:
 a flow which is uniform
 with depth is described by a zero vorticity (irrotational case),
 constant non-zero
 vorticity corresponds to a linear shear flow and non-constant vorticity
 indicates highly sheared flows. See in the Figure 1
 an example of a linear shear flow with
 constant  vorticity $\omega=\textrm{const}:=\omega_0>0$.
 \\

 \hspace{2cm}\scalebox{0.65}{\includegraphics{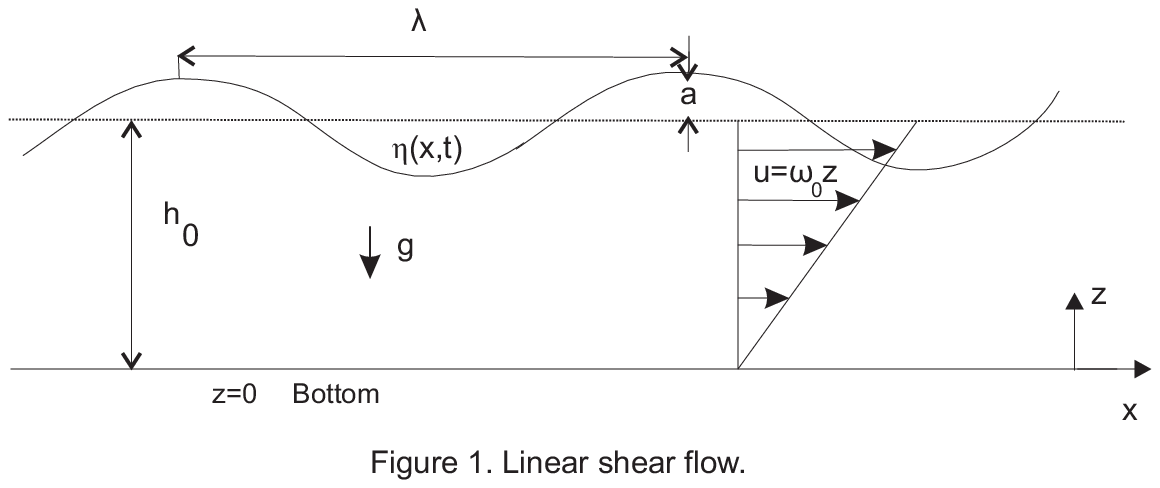}}
\\

 The full Euler equations (\ref{e})-(\ref{bc1'}) are often too complicated
 to analyze directly.
One can pursue for example a mathematical study of their periodic
steady solutions
 in the irrotational case (see see \cite{aft}, \cite{toland}) or
 a study of their  periodic steady
solutions in the case of non-zero vorticity (see \cite{cs2'},
\cite{cs2007}). But in order to reach detailed information about
qualitative features of water waves, it is useful to derive
approximate models which are more amenable to an in-depth
analysis.

\section{Nondimensionalisation and scaling}

In order to develop a systematic approximation procedure, we need
to characterize the water-wave problem (\ref{e})-(\ref{bc1'}) in
terms of the sizes of various fundamental parameters. These
parameters are introduced by defining a set of non-dimensional
variables.\\
First we introduce the appropriate length scales: the undisturbed
depth of water $h_0$, as the vertical scale and a typical
wavelength $\lambda$ (see Figure 1), as the horizontal scale. In
order to define a time scale we require a suitable velocity scale.
An appropriate choice for the scale of the horizontal component of
the velocity is $\sqrt{gh_0}$. Then, the corresponding time scale
is $\f\lambda{\sqrt{gh_0}}$ and the scale for the vertical
component of the velocity is $h_0\f{\sqrt{gh_0}}{\lambda}$. The
surface wave itself leads to the introduction of a typical
amplitude of the wave $a$ (see Figure 1). For more details see
\cite{johnson-carte}. Thus, we define the set of non-dimensional
variables
\begin{equation}
\begin{array}{c}
x\mapsto\lambda x,  \quad z\mapsto h_0 z, \quad \eta\mapsto a\eta,
\quad t\mapsto\f\lambda{\sqrt{gh_0}}t,\\
  u\mapsto  \sqrt{gh_0}u,
\quad v\mapsto h_0\f{\sqrt{gh_0}}{\lambda}v
\end{array} \label{nondim}\end{equation}
where, to avoid new notations, we have used the same symbols for
the non-dimensional variables  $x$, $z$, $\eta$, $t$, $u$, $v$, on
the right-hand side. The partial derivatives will be replaced by
\begin{equation}
\begin{array}{c}
u_t\mapsto \f{gh_0}{\lambda}u_t, \quad u_x\mapsto
\f{\sqrt{gh_0}}{\lambda}u_x, \quad u_z\mapsto\f {\sqrt{gh_0}}{h_0}u_z,\\

v_t\mapsto \f{gh_0^2}{\lambda^2}v_t, \quad v_x\mapsto
h_0\f{\sqrt{gh_0}}{\lambda^2}v_x, \quad v_z\mapsto\f {\sqrt{gh_0}}{\lambda}v_z\\
\end{array}\label{derivate}\end{equation}

\noindent Let us now define the non-dimensional pressure. If the
water would be stationary, that is, $u\equiv v \equiv 0$, from the
equations (\ref{e}) and (\ref{bc2}) with $\eta=0$, we get for a
non-dimensionalised $z$, the hydrostatic pressure $p_0+\rho g
h_0(1-z)$. Thus, the non-dimensional  pressure is defined  by
\begin{equation} p\mapsto p_0+\rho g h_0(1-z)+\rho g h_0 p
\label{p}\end{equation} therefore
\begin{equation} p_x\mapsto \rho \f {gh_0}{\lambda} p_x, \quad
p_z\mapsto -\rho g+\rho g p_z\label{p'}\end{equation}

 Taking
into account (\ref{nondim}), (\ref{derivate}), (\ref{p}) and
(\ref{p'}), the water-wave problem (\ref{e})-(\ref{bc1'})  writes
 in
non-dimensional variables, as \begin{equation}
\begin{array}{c}
u_t+uu_x+vu_z=- p_x\\  \delta^2(v_t+uv_x+vv_z)=- p_z\\
 u_x+v_z=0\\
v=\epsilon(\eta_t+u\eta_x) \,   \textrm{ and } \,  p=\epsilon\eta
\, \, \textrm{ on }\,
z=1+\epsilon\eta(x,t)\\
 v=0 \, \,
\textrm { on } z=0
 \end{array}
\label{e+bc'} \end{equation}  where we have introduced the
amplitude parameter $\epsilon=\f a{h_0}$ and the shallowness
parameter $\delta=\f {h_0}{\lambda}$. In view of (\ref{derivate}),
the
 vorticity equation (\ref{vorticity}) writes in non-dimensional variables
 as
 \be
 u_z=\delta^2v_x+\f{\sqrt{gh_0}}{g}\omega(x,z)\label{vor}
 \ee
For zero vorticity flows (irrotational flows) this equation writes
as
 \be
 u_z=\delta^2v_x\label{vor1}
 \ee
For constant non-zero vorticity flows, that is,
$\omega(x,z)=$const:=$\omega_0$, the equation (\ref{vor}) becomes
\be
 u_z=\delta^2v_x+\f{\sqrt{gh_0}}{g}\omega_0\label{vor2}
 \ee

After the nondimensionalisation of  the system
(\ref{e})-(\ref{bc1'}) let us now proceed  with the scaling
transformation. First we observe that, on $z=1+\epsilon\eta$, both
$v$ and $p$ are proportional to $\epsilon$. This is consistent
with the fact that as $\epsilon\rightarrow 0$ we must have
$v\rightarrow 0$ and $p\rightarrow 0$, and it leads to the
following scaling of the non-dimensional variables
\begin{equation} p\mapsto \epsilon p,\quad
(u,v)\mapsto\epsilon(u,v) \label{scaling}\end{equation} where we
avoided again the introduction of a new notation. The problem
(\ref{e+bc'}) becomes \begin{equation}
\begin{array}{c}
u_t+\epsilon(uu_x+vu_z)=- p_x\\  \delta^2[v_t+\epsilon(uv_x+vv_z)]=- p_z\\
 u_x+v_z=0\\
  v=\eta_t+\epsilon u\eta_x  \, \textrm{ and } \,  p=\eta \, \, \textrm{ on }\,
z=1+\epsilon\eta(x,t)\\
 v=0 \, \,
\textrm { on } z=0
 \end{array}
\label{e+bc1''} \end{equation} and the equation (\ref{vor}) keeps the same form. \\
In what follows we will consider in turn the cases of an
irrotational flow and a constant vorticity flow. The system which
describes our problem  in the irrotational case is given by \be
\begin{array}{c}
u_t+\epsilon(uu_x+vu_z)=- p_x\\  \delta^2[v_t+\epsilon(uv_x+vv_z)]=- p_z\\
 u_x+v_z=0\\
  u_z=\delta^2v_x\\
  v=\eta_t+\epsilon u\eta_x  \, \textrm{ and } \,  p=\eta \, \, \textrm{ on }\,
z=1+\epsilon\eta(x,t)\\
 v=0 \, \,
\textrm { on } z=0
 \end{array}
\label{e+bc''} \end{equation} In the constant vorticity case, the
problem is described  by the following system \be
\begin{array}{c}
u_t+\epsilon(uu_x+vu_z)=- p_x\\  \delta^2[v_t+\epsilon(uv_x+vv_z)]=- p_z\\
 u_x+v_z=0\\
  u_z=\delta^2v_x+\f{\sqrt{gh_0}}{g}\omega_0\\
  v=\eta_t+\epsilon u\eta_x  \, \textrm{ and } \,  p=\eta \, \, \textrm{ on }\,
z=1+\epsilon\eta(x,t)\\
 v=0 \, \,
\textrm { on } z=0
 \end{array}
\label{e+bc2} \end{equation}

\section{The linearized problem}

The two important  parameters $\epsilon$ and $\delta$  that arise
in water-waves theories, are used to define various approximations
of the governing equations and the boundary conditions. The scaled
versions (\ref{e+bc''}) and  (\ref{e+bc2}) of the equations for
our problem, allow immediately the identification of the
linearized problem, by letting $\epsilon\rightarrow 0$, for
arbitrary $\delta$.  The linearized problem in the shallow water
regime is obtain by letting further $\delta\rightarrow 0$. Thus,
we get  the following linear systems, in the irrotational case
\begin{equation}
\begin{array}{c}
u_t+p_x=0\\  p_z=0\\
 u_x+v_z=0\\
 u_z=0\\
v=\eta_t \,  \textrm{ and } \,  p=\eta \, \, \textrm{ on }\,
z=1\\
 v=0 \, \,
\textrm { on } z=0
\end{array}
\label{small} \end{equation} and in the constant vorticity case
\begin{equation}
\begin{array}{c}
u_t+p_x=0\\  p_z=0\\
 u_x+v_z=0\\
 u_z=\f{\sqrt{gh_0}}{g}\omega_0\\
v=\eta_t \,  \textrm{ and } \,  p=\eta \, \, \textrm{ on }\,
z=1\\
 v=0 \, \,
\textrm { on } z=0
\end{array}
\label{small2} \end{equation} From the second equation in
(\ref{small}), respectively (\ref{small2}), we get in the both
cases that $p$ does not depend on $z$. Because $p=\eta(x,t)$ on
$z=1$, we have
\begin{equation} p=\eta(x,t) \, \quad \textrm{ for any } \, \,
0\leq z\leq 1\label{2}\end{equation} Therefore, using the first
equation and the fourth equation in (\ref{small}), respectively
(\ref{small2}), we obtain, in the irrotational case
 \begin{equation} u=-\int_0^t \eta_x(x,s)ds+\mathcal{F}(x)
\label{3''}\end{equation} and in the constant vorticity case
 \be u=-\int_0^t
\eta_x(x,s)ds+\mathcal{F}(x)+\f{\omega_0\sqrt{gh_0}}{g}z\label{32''}\ee
where $\mathcal{F}$ is an arbitrary  function such that
\be
 \mathcal{F}(x)=u(x,0,0)
\ee
\\
 Differentiating
(\ref{3''}), respectively (\ref{32''}), with respect to $x$ and
using the third equation in (\ref{small}), respectively
(\ref{small2}), we get, after an integration against $z$, the same
equation in the both cases \be
v=-zu_x=z\left(\int_0^t\eta_{xx}(x,s)ds
-\mathcal{F}'(x)\right)\label{4}\ee In view of the fifth equation
in (\ref{small}), (\ref{small2}), we get after a differentiation
with respect to $t$, that $\eta$ has to satisfy the equation
 \begin{equation}
\eta_{tt}-\eta_{xx}=0 \label{eta}\end{equation} The general
solution of this equation is $\eta(x,t)=f(x-t)+g(x+t)$, where $f$
and $g$ are differentiable functions.  It is convenient first to
restrict ourselves to waves which propagate in only one direction,
thus, we choose
\begin{equation} \eta(x,t)=f(x-t) \label{sol}\end{equation}
From (\ref{4}), (\ref{sol}) and  the condition $v=\eta_t$ on
$z=1$, we obtain \be \mathcal{F}(x)=f(x)+c_0 \label{c}\ee where
$c_0$ is constant.\\
 Therefore, in the irrotational case, taking
into account (\ref{2}), (\ref{3''}), (\ref{4}), (\ref{sol}) and
(\ref{c}),  the solution of the linear system (\ref{small}) is
given by \be
 \begin{array}{llll}
 \eta(x,t)=f(x-t)\\
 p(x,t)=f(x-t)\\
u(x,z,t)=f(x-t)+c_0\\
v(x,z,t)=-zf'(x-t)=-zu_x \end{array}\label{solrot0}\ee
 In the
case of a constant vorticity flow, from (\ref{2}), (\ref{32''}),
(\ref{4}), (\ref{sol}) and (\ref{c}),
  the solution of the linear system (\ref{small2}) is given by
 \be
 \begin{array}{llll}
 \eta(x,t)=f(x-t)\\
 p(x,t)=f(x-t)\\
u(x,z,t)=f(x-t)+\f{\omega_0\sqrt{gh_0}}{g}z+c_0\\
v(x,z,t)=-zf'(x-t)=-zu_x \end{array}\label{solrotconst}\ee

\section{Particle trajectories beneath linearized shallow water waves}

Let $\left(x(t), z(t)\right)$ be the path of a particle in the
fluid domain, with location $\left(x(0), z(0)\right)$ at time
$t=0$. The motion of the particle is described by the differential
system \be
 \left\{\begin{array}{ll}
 \f{dx}{dt}=u(x,z,t)\\
 \f{dz}{dt}=v(x,z,t)
 \end{array}\right.\label{diff}\ee
with the initial data $\left(x(0), z(0)\right):=(x_0,z_0)$.

\subsection{The case of an irrotational flow}

This case was investigated in \cite{io}. We summarize below the
results presented in detail in reference \cite{io}. Some new
observations concerning the solutions shown in  Fig. 2 and Fig. 6
are added.

 Making the
\textit{Ansatz}

\be f(x-t)=\cos(2\pi(x-t))\label{ansatz} \ee from (\ref{solrot0}),
the differential system (\ref{diff}) becomes
\be\left\{\begin{array}{ll}
 \f{dx}{dt}=\cos(2\pi(x-t))+c_0\\
 \\
 \f{dz}{dt}=2\pi z\sin(2\pi(x-t))
 \end{array}\right.\label{diff2}\ee
Notice that the constant $c_0$ is the average of the horizontal
fluid
 velocity over any horizontal
 segment of length 1, that is,
 \be
 c_0=\f 1 {1}\int_{x}^{x+1}u(s,z,t)ds,
 \ee
 representing therefore the strength of the underlying uniform
 current. Thus, $c_0=0$  will correspond to a region of still water with
 no underlying current,
 $c_0>0$ will characterize a favorable uniform current and  $c_0<0$
 will characterize an adverse uniform current.

 The right-hand side of the differential system (\ref{diff2})
 is smooth and bounded, therefore, the unique solution of the Cauchy
 problem with initial data $(x_0,z_0)$ is defined globally in
 time.

 To study the exact solution of the system (\ref{diff2}) it is
 more convenient to re-write it in the following moving frame
 \be
 X=2\pi(x-t),\quad  Z=z \label{frame}
 \ee
This transformation yields \be\left\{\begin{array}{ll}
 \f{dX}{dt}=2\pi\cos(X)+2\pi(c_0-1)\\
 \\
 \f{dZ}{dt}=2\pi Z\sin(X)
 \end{array}\right.\label{diff3}\ee
For $\mathbf{c_0=0}$,  we get the following  exact solution of the
system (\ref{diff3}):
 \be
X(t)=2\textrm{arccot }(2\pi t+a), \quad a =\textrm{ constant
}\label{X}\ee \be Z(t)=Z(0)\exp \Bigg(\int_0^t2\pi \sin(X(s))\,
ds\Bigg) =Z(0)\exp\Bigg(\ln\Big[\f{1+(2\pi
t+a)^2}{1+a^2}\Big]\Bigg)\label{Z} \ee Taking into account
(\ref{frame}), (\ref{X}) and (\ref{Z}), we obtain the solution of
the system (\ref{diff2}) with the initial data $(x_0,z_0)$:
\be\left\{
\begin{array}{ll}x(t)=t+\f
1{\pi}\textrm{arccot }(2\pi t+a)\\
\cr z(t)=\f{z_0}{1+a^2}[1+(2\pi t+a)^2]
\end{array}\right.\label{solutie}\ee   From the initial conditions,
we have $a:=\textrm{cot }(\pi x_0)$.

\noindent Studying  the derivatives of $x(t)$, $z(t)$ with respect
to $t$ and the limits of $x(t)$, $z(t)$, $\f{z(t)}{x(t)}$ for
$t\rightarrow \pm \infty$,  we draw the graph of the parametric
curve (\ref{solutie}):

\vspace{0.6cm}

\hspace{2cm}\scalebox{0.70}{\includegraphics{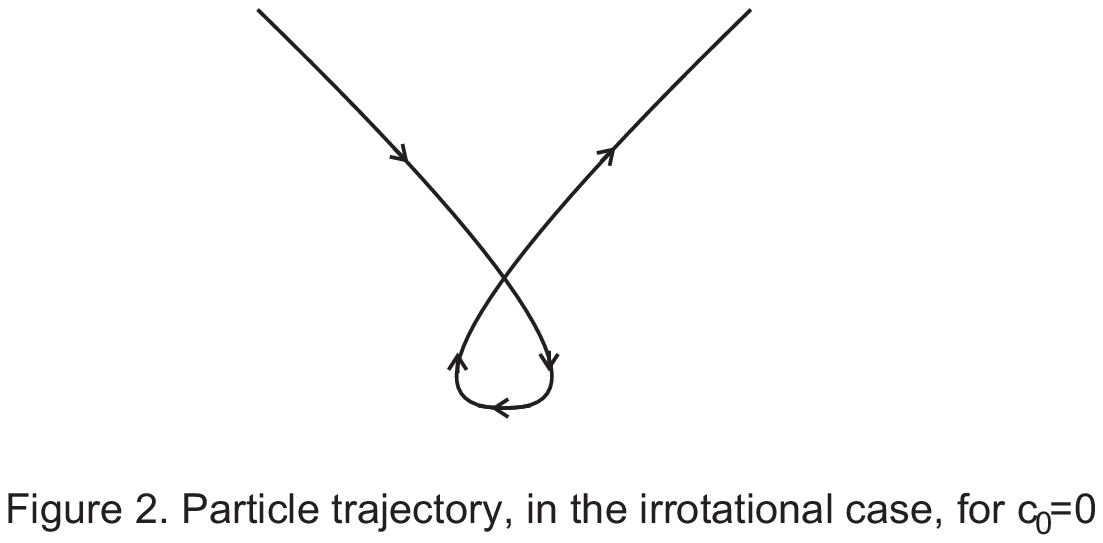}}
\\

\noindent As is shown in Figure 2, the coordinate $z$ increases
indefinitely in time, thus, the solution (\ref{solutie}) is
\textit{not} physically acceptable. In dealing with the linearized
problem not with the full governing equations, we expect to appear
solutions which are not physically acceptable.
\\

\noindent For $\mathbf{c_0(c_0-2)>0}$, the exact solution of the
system (\ref{diff3}) has the following expression:
 \be X(t)=2\textrm{arccot
}\Big[\mathfrak{C}_0\tan\left(\alpha(t)\right)\Big], \label{X1}\ee
\ba Z(t)=Z(0)\exp\Bigg(\int_0^t
\f{4\pi\mathfrak{C}_0\tan\left(\alpha(s)\right)}{1+\Big[\mathfrak{C}_0\tan
\left(\alpha(s))\right)\Big]^2}ds\Bigg) \label{Z'} \ea where \be
\mathfrak{C}_0:=\sqrt{\f{c_0-2}{c_0}} \ee \be
\alpha(t):=-\f{c_0\mathfrak{C}_0}{2}(2\pi t+a)\ee
  From
(\ref{frame}), (\ref{X1}) and (\ref{Z'}), we obtain  the solution
of the system (\ref{diff2}) with the initial data $(x_0,z_0)$,
$z_0>0$,
 \be\left\{
\begin{array}{ll}x(t)=t+\f
1{\pi}\textrm{arccot
}\Big[\mathfrak{C}_0\tan\left(\alpha(t)\right)\Big]
\\
\cr z(t)=z_0 \exp\Bigg( \int_0^t
\f{4\pi\mathfrak{C}_0\tan\left(\alpha(s)\right)}{1+\Big[\mathfrak{C}_0\tan
\left(\alpha(s))\right)\Big]^2} ds\Bigg)\end{array}\right.\ee
Studying the derivatives of $x(t)$ and $z(t)$ with respect to $t$,
we get the following graphs:

\vspace{1cm}

\hspace{2cm}\scalebox{0.70}{\includegraphics{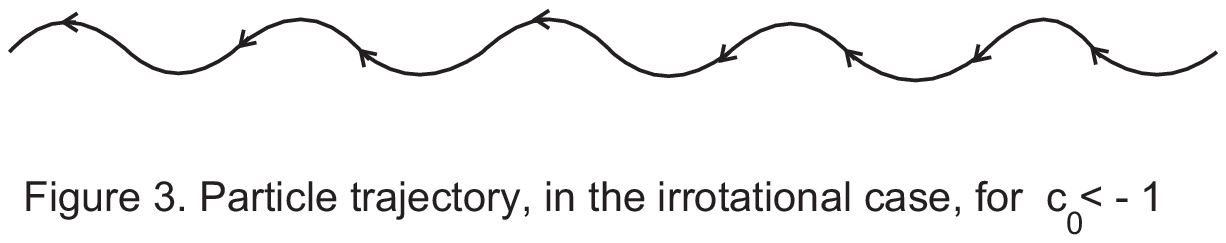}}
\\

\vspace{1cm}

\hspace{2cm}\scalebox{0.70}{\includegraphics{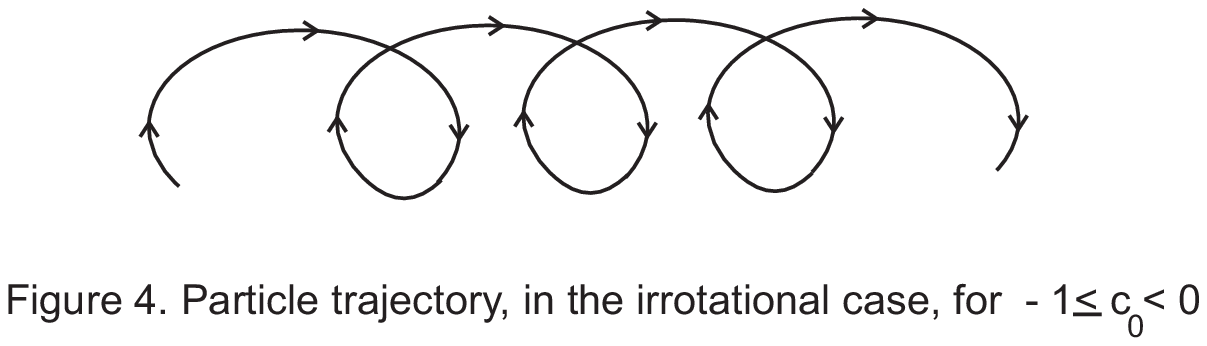}}
\\

\vspace{1cm}

\hspace{2cm}\scalebox{0.70}{\includegraphics{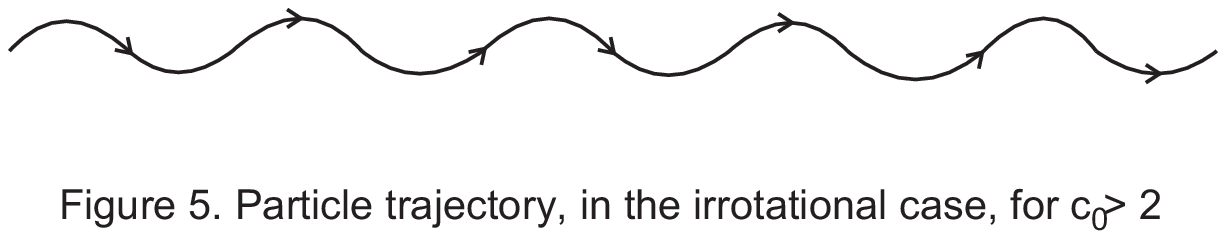}}
\\

\noindent For $\mathbf{c_0\in (0,2]}$, the exact solution of the
system (\ref{diff3}) has for
$|\cot\left(\f{X}{2}\right)|>\mathfrak{K}_0$ the following
expression: \be X(t)=2\textrm{arccot
}\Big[\mathfrak{K}_0\coth\left(\beta(t)\right)\Big], \quad
Z(t)=Z(0)\exp\Bigg(\int_0^t
\f{4\pi\mathfrak{K}_0\coth\left(\beta(s)\right)}{1+\Big[\mathfrak{K}_0\coth
\left(\beta(s))\right)\Big]^2}ds\Bigg),\label{X1'}\ee and for
$|\cot\left(\f{X}{2}\right)|<\mathfrak{K}_0$ the following
expression: \be X(t)=2\textrm{arccot
}\Big[\mathfrak{K}_0\tanh\left(\beta(t)\right)\Big], \quad
Z(t)=Z(0)\exp\Bigg(\int_0^t
\f{4\pi\mathfrak{K}_0\tanh\left(\beta(s)\right)}{1+\Big[\mathfrak{K}_0\tanh
\left(\beta(s))\right)\Big]^2}ds\Bigg)\label{X1''}\ee  In the
above formulas  we have denoted by \be
\mathfrak{K}_0:=\sqrt{\f{2-c_0}{c_0}} \ee \be
\beta(t):=\f{c_0\mathfrak{K}_0}{2}(2\pi t+a)\ee
 From
(\ref{frame}), (\ref{X1'}) and (\ref{X1''}), we obtain  the
solution of the system (\ref{diff2}) with the initial data
$(x_0,z_0)$, $z_0>0$, \be\left\{
\begin{array}{ll}x(t)=t+\f
1{\pi}\textrm{arccot
}\Big[\mathfrak{K}_0\coth\left(\beta(t)\right)\Big]
\\
\cr z(t)=z_0 \exp\Bigg( \int_0^t
\f{4\pi\mathfrak{K}_0\coth\left(\beta(s)\right)}{1+\Big[\mathfrak{K}_0\coth
\left(\beta(s))\right)\Big]^2} ds\Bigg)
\end{array}\right.\label{sol1}\ee or \be\left\{
\begin{array}{ll}x(t)=t+\f
1{\pi}\textrm{arccot
}\Big[\mathfrak{K}_0\tanh\left(\beta(t)\right)\Big]
\\
\cr z(t)=z_0 \exp\Bigg( \int_0^t
\f{4\pi\mathfrak{K}_0\tanh\left(\beta(s)\right)}{1+\Big[\mathfrak{K}_0\tanh
\left(\beta(s))\right)\Big]^2} ds\Bigg)
\end{array}\right.\label{sol2}\ee

\noindent The graphs of the parametric curves in (\ref{sol1}),
(\ref{sol2}) are drawn in figure below
 \vspace{0.7cm}

\hspace{1cm}\scalebox{0.70}{\includegraphics{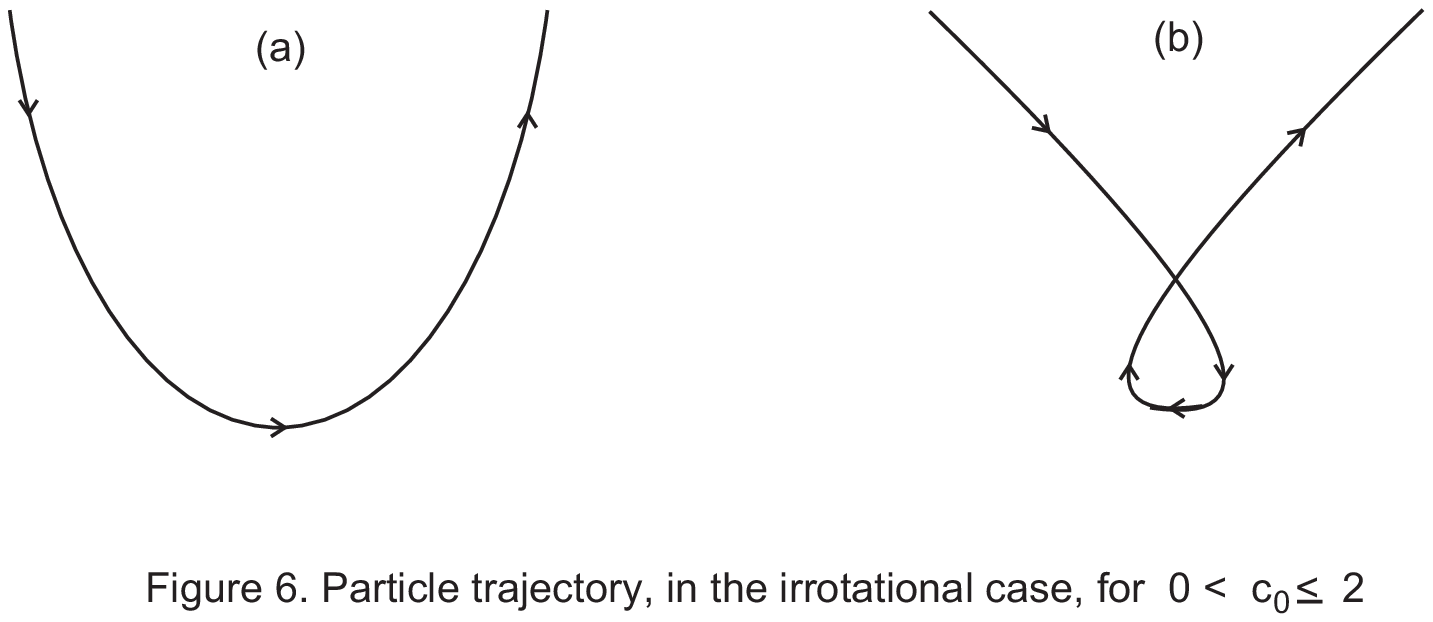}}
\\

\noindent As is shown in Figure 6, the coordinate $z$ increases
indefinitely in time, hence, the solutions (\ref{sol1}),
(\ref{sol2}), are \textit{not} physically acceptable.

\noindent Thus, one gets the following theorem:


\vspace{0.15cm}

\textsc{Theorem 1.}

 \textit{
In the case that the underlying uniform current is moving in the
same direction as an irrotational shallow water wave and the
strength of the current is bigger than 2, then  the particle
trajectories beneath the wave are undulating paths to the right
(see Figure 5).}

\textit{In the case that the underlying uniform current is moving
in the opposite  direction as an irrotational shallow water wave
and the strength of the current is smaller than -1, then  the
particle trajectories beneath the wave are undulating paths to the
left (see  Figure 3). If the strength of the adverse current is
bigger than -1, then the particle trajectories are loops with
forward drift (see Figure 4).}

\textit{In the case of no underlying current and in  the case that
the underlying uniform current is moving in the same direction as
an irrotational shallow water wave with the strength of the
current smaller than 2, the particle trajectories obtained (see
Figure 2 and Figure 6) are not physically acceptable. In these
cases it seems necessary to study the full nonlinear problem.}

\vspace{0.15cm}

\subsection{The case of a constant vorticity flow}

Making the \textit{Ansatz}

\be f(x-t)=\cos(2\pi(x-t))\label{ansatz1} \ee from
(\ref{solrotconst}), the differential system (\ref{diff}) becomes
\be\left\{\begin{array}{ll}
 \f{dx}{dt}=\cos(2\pi(x-t))+\f{\omega_0\sqrt{gh_0}}{g}z+c_0\\
 \\
 \f{dz}{dt}=2\pi z\sin(2\pi(x-t))
 \end{array}\right.\label{diff2'}\ee
 Notice that the constant $c_0$ is the average of the horizontal
fluid
 velocity on the bottom over any horizontal
 segment of length 1, that is,
 \be
 c_0=\f 1 {1}\int_{x}^{x+1}u(s,0,t)ds,
 \ee

The right-hand side of the differential system (\ref{diff2'})
 is smooth and bounded, therefore, the unique solution of the
Cauchy
 problem with initial data $(x_0,z_0)$ is defined globally in
 time.

 To study the exact solution of the system (\ref{diff2'}) it is
 more convenient to re-write it in the following moving frame
 \be
 X=2\pi(x-t),\quad  Z=z \label{frame'}
 \ee
This transformation yields \be\left\{\begin{array}{ll}
 \f{dX}{dt}=2\pi\cos(X)+2\pi\f{\omega_0\sqrt{gh_0}}{g}Z +2\pi(c_0-1)\\
 \\
 \f{dZ}{dt}=2\pi Z\sin(X)
 \end{array}\right.\label{diff3'}\ee
Let us now investigate the differential system (\ref{diff3'})\\
Differentiating with respect to $t$ the first equation in
(\ref{diff3'}) and taking into account  (\ref{diff3'}), this first
equation becomes \be \f{d^2 X}{dt^2}=4\pi^2 \sin
(X)\left[1-c_0-\cos(X)\right] \label{ecX}\ee
 Like
in the irrotational case (see \cite{io}), we use the following
substitution (see \cite{kamke}, I.76, page 308) \be
\cot\left(\f{X}{2}\right)=y\, ,\quad \sin(X)=\f{2y}{y^2+1}\,
,\quad \cos(X)=\f{y^2-1}{y^2+1}\,, \quad
dX=-\f{2}{y^2+1}dy\label{substitution}\ee In the new variable, the
 equation (\ref{ecX}) takes the form
 \be
 \f{d^2 y}{dt^2}-\f{2y}{y^2+1}\left(\f{dy}{dt}\right)^2+\f{8\pi^2
 y}{y^2+1}-4\pi^2c_0 y=0\label{ecy}\ee
 The  solution of the
system (\ref{diff3'}) has then the following expression:
 \be
X(t)=2\textrm{arccot }(y(t)), \label{Xy}\ee \be Z(t)=Z(0)\exp
\Bigg(\int_0^t\f{4\pi y(s)}{y^2(s)+1} \, ds\Bigg)\label{Zy} \ee
with $y(t)$ satisfying the ordinary differential equation
(\ref{ecy}).\\
 From (\ref{frame'}), (\ref{Xy}) and (\ref{Zy}), we
obtain the solution of the system (\ref{diff2'})  with the initial
data $(x_0,z_0)$:  \be\left\{
\begin{array}{ll}x(t)=t+\f
1{\pi}\textrm{arccot }(y(t))\\
\cr z(t)=z_0\exp \Bigg(\int_0^t\f{4\pi y(s)}{y^2(s)+1} \, ds\Bigg)
\end{array}\right.\label{solutie'}\ee
where $y(t)$ satisfies the ordinary differential equation
(\ref{ecy}) with the following initial conditions: \ba
&&y(0)=\cot(\pi x_0)\nonumber\\
&&\f{dy}{dt}(0)=\pi\Bigg[2-\Big(\f{\omega_0\sqrt{gh_0}}{g}z_0+c_0\Big)
\Big(\cot^2(\pi x_0)+1\Big)\Bigg] \label{16}\ea obtained from
(\ref{substitution}) and the first equation in (\ref{diff3'}).

\subsubsection{The case $\mathbf{c_0=0}$}

In this case the equation (\ref{ecy}) writes as \be\f{d^2
y}{dt^2}=\f{2y}{y^2+1}\left[\left(\f{dy}{dt}\right)^2-4\pi^2\right]\label{5}\ee
For \be\f{dy}{dt}\neq \pm 2\pi\, \Longleftrightarrow\, y\neq \pm
2\pi t+a, \, \textrm{ with } \, a\,  \textrm{ constant },\ee we
put the equation (\ref{5}) into the following form \be
\f{y'}{(y')^2-4\pi^2}y''=\f{2y}{y^2+1}y',\label{6}
\ee where $y':=\f{dy}{dt},$ $y'':=\f{d^2y}{dt^2}$.\\
 We
integrate in (\ref{6}) and we get \be (y')^2-4\pi^2=A^2
(y^2+1)^2\, \Longleftrightarrow \, (y')^2=A^2(y^2+1)^2+4\pi^2,
\label{7}\ee $A$ being an integration constant, which, taking into
account (\ref{16}), can be expressed function of the initial data
$(x_0, z_0)$ and of the constant vorticity $\omega_0$.\\
 The
solution of the equation (\ref{7}) involves an elliptic integral
of first kind, that is, \be
\pm\int\f{dy}{\sqrt{A^2(y^2+1)^2+4\pi^2}}=t \label{8}\ee The
elliptic integral of first kind  from (\ref{8}) may by reduced to
Legendre's normal form. In order to do this we first consider the
substitution \be y^2=w \label{11}\ee The left hand side in
(\ref{8}) becomes
 \ba \pm\int\f{dy}{\sqrt{A^2(y^2+1)^2+4\pi^2}}=\pm\f{1}{2|A|}\int\f{dw}{\sqrt{w\left(w^2+2w+1+\f{4\pi^2}{A^2}\right)
}}\nonumber \ea We introduce now the variable $\varphi$ by (see
\cite{smirnov} Ch. VI, \S 4, page 603) \be
w=\sqrt{1+\f{4\pi^2}{A^2}}\tan^2\f{\varphi}{2} \label{12}\ee
 and we get
\be
w\left(w^2+2w+1+\f{4\pi^2}{A^2}\right)=\left(\sqrt{1+\f{4\pi^2}{A^2}}\right)^3\left(1-k^2\sin^2\varphi\right)
\f{\tan^2\f{\varphi}{2}}{\cos^4\f{\varphi}{2}} \nonumber\ee \be
dw= \sqrt{1+\f{4\pi^2}{A^2}}\f{\tan\f{\varphi}{2}}{\cos^2
\f{\varphi}{2}}d\varphi\nonumber\ee where the constant $0<k^2<1$
is given by \be
k^2=\f{1}{2}\left(1-\f{1}{\sqrt{1+\f{4\pi^2}{A^2}}}\right)\nonumber
\ee Therefore we obtain the Legendre normal form of the integral
in (\ref{8}), that is, \be \pm
\f{1}{2|A|\left(1+\f{4\pi^2}{A^2}\right)}\int
\f{d\varphi}{\sqrt{1-k^2\sin^2\varphi}}.\label{legendre1}\ee

  Taking into account (\ref{7}), the derivatives of $x(t)$ and
$z(t)$ from (\ref{solutie'}) with respect to $t$, have the
expressions \be
\begin{array}{ll}
x'(t)=\f{\pi(y^2+1)\mp \sqrt{A^2(y^2+1)^2+4\pi^2}}{\pi(y^2+1)}\\
\cr z'(t)= \f{4\pi z_0y}{y^2+1}\exp \Bigg(\int_0^t\f{4\pi
y(s)}{y^2(s)+1} \, ds\Bigg)
\end{array}\label{9}\ee
where $y=y(t)$ is given implicitly  by (\ref{8}).

\noindent For the alternative with "+" in the expression (\ref{9})
of $x'(t)$, we obtain that $x'(t)>0$, for all $t$. The sign of
$z'(t)$
 depends on  $y(t)$.  For $y(t)<0$ we get $z'(t)<0$, and for
$y(t)>0$ we get $z'(t)>0.$ Concerning the zeros of the function
$y(t)$ we can tell more looking at the Legendre normal form of the
integral in (\ref{8}), that is, at the expression
(\ref{legendre1}). The inverse of this integral  is, excepting a
constant factor, the Jacobian elliptic function
sn($t$):=$\sin\varphi$ (see, for example, \cite{byrd}). Thus,
taking into account (\ref{11}), (\ref{12}) and the formula
$\tan\f{\varphi}{2}=\f{\sin \varphi}{1+\cos\varphi}$, we get that
the zeros of the function $y(t)$ are the zeros of the periodic
Jacobian elliptic function sn$(t)$ (for the graph of sn$(t)$ see,
for example, \cite{byrd}, Figure 10, page 26).
 Therefore, the particle trajectory
looks like in Figure 7 (a).

\noindent Let us consider now  the alternative with "-" in the
expression
(\ref{9}) of $x'(t)$.\\
If \be |A|>\pi \ee then we have $x'(t)<0$, for all $t$. The sign
of $z'(t)$
 depends on  $y(t)$: for $y(t)<0$ we have $z'(t)<0$, and for
$y(t)>0$ we have $z'(t)>0.$  The function $y(t)$  being periodic
and having multiple zeros, the
particle trajectory looks like in Figure 7 (c).\\
If \be |A|<\pi \ee then the zeros of $x'(t)$ are obtained for \be
y(t)=\pm \sqrt{\f{2\pi}{\sqrt{\pi^2-A^2}}-1}\ee
 The sign of $z'(t)$  depends on  $y(t)$ as above, that is,
 $z'(t)<0$ for $y(t)<0$, and $z'(t)>0$ for
$y(t)>0$. Summing up,\\
for $y(t)\in (-\infty, -\sqrt{\f{2\pi}{\sqrt{\pi^2-A^2}}-1})$ we
have $x'(t)>0$, $z'(t)<0$,\\
 for $y(t)\in
(-\sqrt{\f{2\pi}{\sqrt{\pi^2-A^2}}-1},0)$ we have $x'(t)<0$,
$z'(t)<0$,\\
 for $y(t)\in (0,
\sqrt{\f{2\pi}{\sqrt{\pi^2-A^2}}-1})$ we have $x'(t)<0$,
$z'(t)>0$, \\
for $y(t)\in (\sqrt{\f{2\pi}{\sqrt{\pi^2-A^2}}-1}, \infty)$ we
have $x'(t)>0$, $z'(t)>0$.\\
If the periodic function y(t) has values in each interval from
above, then the particle trajectory looks like in Figure 7 (b).

\vspace{0.15cm}

\textsc{Theorem 2.}

\textit{ As periodic waves propagate on the water's free surface
of a constant vorticity shallow water flow over a flat bed,  with
the average of the horizontal fluid velocity on the bottom over
any horizontal segment of length 1 equals zero, there does not
exist only a single pattern for all the water particles. The
particle paths are not closed, and depending on the relation
between the initial data $(x_0,z_0)$ and the constant vorticity
$\omega_0$, some particle trajectories are undulating curves to
the right (see Figure 7 (a)), or to the left (see Figure 7 (c)),
others are loops with forward drift (see Figure 7 (b)).}

\vspace{0.15cm}

\subsubsection{The case $\mathbf{c_0\neq 0}$}

In this case we transform the equation (\ref{ecy}), by (see
\cite{kamke}, 6.45, page 551)
 \be
w(y):=\left(\f{dy}{dt}\right)^2,\label{w} \ee into the following
equation \be \f 1{2}\f{dw}{dy}-\f{2y}{y^2+1}w(y)+\f{8\pi^2
 y}{y^2+1}-4\pi^2c_0 y=0 \label{ecy'}\ee
 The homogeneous equation:
 \be
 \f 1{2}\f{dw}{dy}-\f{2y}{y^2+1}w(y)=0
 \ee
has the solution \be w_h(y)=B(y^2+1)^2\ee where $B$ is an
integration constant. By the method of variation of constants, the
general solution of the non-homogeneous equation (\ref{ecy'}) is
given by \be w(y)=B(y)(y^2+1)^2 \ee where $B(y)$ is a continuous
function which satisfies the equation \be \f{dB}{dy}=
\f{8\pi^2c_0y}{(y^2+1)^2}-\f{16\pi^2 y}{(y^2+1)^3} \label{10}\ee
The solution of the equation (\ref{10}) is \be
B(y)=-\f{4\pi^2c_0}{y^2+1}+\f{4\pi^2}{(y^2+1)^2} +\mathcal{C}\ee
where $\mathcal{C}$ is constant. Therefore, the solution of the
non-homogeneous equation (\ref{ecy'}) has the expression \be
w(y)=\mathcal{C}(y^2+1)^2 -4\pi^2c_0(y^2+1)+4\pi^2\ee Taking into
account (\ref{w}), we now obtain, instead of the equation
(\ref{7}), the following equation \be (y')^2= \mathcal{C}(y^2+1)^2
-4\pi^2c_0(y^2+1)+4\pi^2\label{7'}\ee where $y':=\f{dy}{dt}$, and
the following condition has  to be satisfied \be
\pi^2c_0^2<\mathcal{C}\label{C}\ee in order to have the right hand
side in
(\ref{7'}) bigger then zero for any $y$.\\
 The solution of the equation
(\ref{7'}) involves an elliptic integral of first kind \be
\pm\int\f{dy}{\sqrt{\mathcal{C}(y^2+1)^2
-4\pi^2c_0(y^2+1)+4\pi^2}}=t, \label{8'} \ee The elliptic integral
of first kind  from (\ref{8'}) may by reduced to Legendre's normal
form. In order to do this we first consider the substitution \be
y^2=w \nonumber\ee The left hand side in (\ref{8'}) becomes
 \ba &&
\pm\int\f{dy}{\sqrt{\mathcal{C}(y^2+1)^2
-4\pi^2c_0(y^2+1)+4\pi^2}}\nonumber\\
=&&\pm\f{1}{2}\int\f{dw}{\sqrt{\mathcal{C}w\left[w^2+w\left(2-\f{4\pi^2
c_0}{\mathcal{C}}\right)+1+\f{4\pi^2}{\mathcal{C}}-\f{4\pi^2c_0}{\mathcal{C}}\right]}}
\nonumber \ea We introduce now the variable $\varphi$ by (see
\cite{smirnov} Ch. VI, \S 4, page 603) \be
w=\sqrt{1+\f{4\pi^2}{\mathcal{C}}-\f{4\pi^2c_0}{\mathcal{C}}}\tan^2\f{\varphi}{2}
\nonumber\ee
 and we get
\be w\left[w^2+w\left(2-\f{4\pi^2
c_0}{\mathcal{C}}\right)+1+\f{4\pi^2}{\mathcal{C}}-\f{4\pi^2c_0}{\mathcal{C}}\right]
=\left(\sqrt{1+\f{4\pi^2}{\mathcal{C}}-\f{4\pi^2c_0}{\mathcal{C}}}\right)^3\left(1-k^2\sin^2\varphi\right)
\f{\tan^2\f{\varphi}{2}}{\cos^4\f{\varphi}{2}} \nonumber\ee \be
dw=
\sqrt{1+\f{4\pi^2}{\mathcal{C}}-\f{4\pi^2c_0}{\mathcal{C}}}\f{\tan\f{\varphi}{2}}{\cos^2
\f{\varphi}{2}}d\varphi\nonumber\ee where the constant $0<k^2<1$
is given by \be k^2=\f{1}{2}\left(1-
\f{\mathcal{C}-2\pi^2c_0}{\mathcal{C}\sqrt{1+\f{4\pi^2}{\mathcal{C}}-\f{4\pi^2c_0}{\mathcal{C}}}}\right)\nonumber
\ee Therefore we obtain the Legendre normal form of the integral
in (\ref{8'}), that is, \be \pm
\f{1}{2\sqrt{\mathcal{C}}\left(1+\f{4\pi^2}{\mathcal{C}}-\f{4\pi^2c_0}{\mathcal{C}}
\right)}\int
\f{d\varphi}{\sqrt{1-k^2\sin^2\varphi}}\label{legendre2}\ee

 Taking into account (\ref{7'}), the derivatives of $x(t)$ and
$z(t)$ from (\ref{solutie'}) with respect to $t$, have the
expressions \be
\begin{array}{ll}
x'(t)=\f{\pi(y^2+1)\mp \sqrt{\mathcal{C}(y^2+1)^2
-4\pi^2c_0(y^2+1)+4\pi^2}}{\pi(y^2+1)}\\
\cr z'(t)= \f{4\pi z_0y}{y^2+1}\exp \Bigg(\int_0^t\f{4\pi
y(s)}{y^2(s)+1} \, ds\Bigg)
\end{array}\label{9'}\ee
where $y=y(t)$ is given implicitly  by (\ref{8'}).\\
For the alternative with "+" in the expression (\ref{9'}) of
$x'(t)$, we obtain that $x'(t)>0$, for all $t$. The sign of
$z'(t)$
 depends on  $y(t)$. For $y(t)<0$ we get $z'(t)<0$, and for
$y(t)>0$ we get $z'(t)>0.$ Taking into account the Legendre normal
form of the integral in (\ref{8'}), that is, the expression  in
(\ref{legendre2}), and using the same arguments as in the case
$c_0=0$ on the page 14, we get that the zeros of $y(t)$ are the
zeros of the periodic Jacobian elliptic function sn$(t)$.
Therefore, in this case the particle trajectory looks like in Figure 7 (a).\\
Let us consider now  the alternative with "-" in the expression
(\ref{9'}) of $x'(t)$. The zeros of $x'(t)$ are obtained by
solving the equation \be
(\pi^2-\mathcal{C})(y^2+1)^2+4\pi^2c_0(y^2+1)-4\pi^2=0\label{13}\ee
The discriminant of the quadratic equation in $W$ \be
(\pi^2-\mathcal{C})W^2+ 4\pi^2 c_0 W-4\pi^2=0 \label{W}\ee is \be
\Delta=16\pi^2(\pi^2c_0^2+\pi^2-\mathcal{C}) \label{delta}\ee

 I)
If \be \mathcal{C}>\pi^2c_0^2+\pi^2\ee
 then $\Delta<0$ and $\pi^2-\mathcal{C}<0$. This yields
$x'(t)<0$, for all $t$. The sign of $z'(t)$
 depends on  $y(t)$ as was mentioned above. The function $y(t)$ being periodic
 and having multiple zeros, the  particle trajectory looks like in Figure 7 (c).\\

II) If \be \mathcal{C}<\pi^2c_0^2+\pi^2\label{14}\ee
 then $\Delta>0$ and the equation (\ref{W}) has two real
 solutions.

 II a)
 If in addition to the condition (\ref{14})
we have \be \pi^2-\mathcal{C}>0\ee then, one of the solution of
the equation (\ref{W}) is positive and the other one is negative.
We denote the positive solution of the equation (\ref{W}) by \be
W_1:=\f{-2\pi^2c_0+2\pi\sqrt{\pi^2c_0^2+\pi^2-\mathcal{C}}}{\pi^2-\mathcal{C}}
\ee
 Further, if $W_1-1<0$, then  the sign of $x'(t)$ is given
by $\pi^2-\mathcal{C}$, that is, in this case $x'(t)>0$, for all
$t$. Taking into account that $z'(t)>0$ for $y(t)>0$, and
$z'(t)<0$ for $y(t)<0$, the particle trajectory looks now like in
7(a).\\
 If $W_1-1>0$, then the zeros of $x'(t)$ are obtained for
\be y(t)=\pm\sqrt{W_1-1} \ee
The sign of $z'(t)$  depends on  $y(t)$ as above. Thus,\\
for $y(t)\in (-\infty, -\sqrt{W_1-1})$ we
have $x'(t)>0$, $z'(t)<0$,\\
 for $y(t)\in
(-\sqrt{W_1-1},0)$ we have $x'(t)<0$,
$z'(t)<0$,\\
 for $y(t)\in (0,
\sqrt{W_1-1})$ we have $x'(t)<0$,
$z'(t)>0$, \\
for $y(t)\in (\sqrt{W_1-1}, \infty)$ we
have $x'(t)>0$, $z'(t)>0$.\\
If the periodic function y(t) has values in each interval from
above, then the particle trajectory looks like in Figure 7 (b).\\

 II b)  If in addition to the condition (\ref{14})
we have \be \pi^2-\mathcal{C}<0\label{15}\ee then, the solutions
of the equation (\ref{W}) are both  negative or both positive.

If \be c_0<0 \ee then both solutions of (\ref{W}) are negative.
Therefore, the sign of $x'(t)$ is given by $\pi^2-\mathcal{C}$,
that is, in this case $x'(t)<0$, for all $t$. How $z'(t)<0$ for
$y(t)<0$ and $z'(t)>0$ for $y(t)>0$, the particle trajectory looks
like in 7(c).

 If \be c_0>0 \ee
then both solutions of (\ref{W}) are positive. We denote these
positive solutions by \ba
W_1:=\f{-2\pi^2c_0+2\pi\sqrt{\pi^2c_0^2+\pi^2-\mathcal{C}}}{\pi^2-\mathcal{C}},\nonumber\\
W_2:=\f{-2\pi^2c_0-2\pi\sqrt{\pi^2c_0^2+\pi^2-\mathcal{C}}}{\pi^2-\mathcal{C}}
\label{w1w2}\ea
We observe that  $W_1<W_2$.\\
 Further, if
$W_1-1<0$ and $W_2-1<0$ then the sign of $x'(t)$ is given by
$\pi^2-\mathcal{C}$, that is, in this case $x'(t)<0$, for all $t$.
Therefore, the particle trajectory looks like in 7(c).\\
 If $W_1-1<0$ and $W_2-1>0$ then
the zeros of $x'(t)$ are obtained for \be y(t)=\pm\sqrt{W_2-1} \ee
The sign of $z'(t)$ depends on $y(t)$ as above. Thus, taking into
account
 (\ref{15}), \\
for $y(t)\in (-\infty, -\sqrt{W_2-1})$ we
get $x'(t)<0$, $z'(t)<0$,\\
 for $y(t)\in
(-\sqrt{W_2-1},0)$ we get $x'(t)>0$,
$z'(t)<0$,\\
 for $y(t)\in (0,
\sqrt{W_2-1})$ we get $x'(t)>0$,
$z'(t)>0$, \\
for $y(t)\in (\sqrt{W_2-1}, \infty)$ we
get $x'(t)<0$, $z'(t)>0$.\\
If the periodic function y(t) has values in each interval from
above, then the particle trajectory looks like in Figure 7 (d).\\
If $W_1-1>0$ and $W_2-1>0$ then the zeros of $x'(t)$ are obtained
for \be y(t)=\pm\sqrt{W_1-1},\quad y=\pm \sqrt{W_2-1} \ee The sign
of $z'(t)$  depends on  $y(t)$ as above. Thus, taking into account
 (\ref{15}), \\
for $y(t)\in (-\infty, -\sqrt{W_2-1})$ we
have $x'(t)<0$, $z'(t)<0$,\\
 for $y(t)\in
(-\sqrt{W_2-1},-\sqrt{W_1-1})$ we have $x'(t)>0$,
$z'(t)<0$,\\
for $y(t)\in(-\sqrt{W_1-1},0)$ we have $x'(t)<0$,
$z'(t)<0$,\\
for $y(t)\in (0, \sqrt{W_1-1})$ we have $x'(t)<0$,
$z'(t)>0$, \\
for $y(t)\in (\sqrt{W_1-1},\sqrt{W_2-1})$ we have $x'(t)>0$,
$z'(t)>0$, \\
for $y(t)\in (\sqrt{W_2-1}, \infty)$ we
have $x'(t)<0$, $z'(t)>0$.\\
If the periodic function y(t) has values in each interval from
above, then the particle trajectory  has the peculiar shape  in
Figure 7 (e).



\vspace{0.15cm}

\textsc{Theorem 3.}

\textit{As periodic waves propagate on the water's free surface of
a constant vorticity shallow water flow over a flat bed,  with the
average of the horizontal fluid velocity on the bottom over any
horizontal segment of length 1 different from zero, there does not
exist only a single pattern for all the water particles. The
particle paths are not closed, and depending on the relation
between the initial data $(x_0,z_0)$ and the constant vorticity
$\omega_0$,  some particle trajectories are undulating curves to
the right (see Figure 7 (a)), or to the left (see Figure 7 (c)),
others are loops with forward drift (see Figure 7 (b)), or with
backward drift (See Figure 7 (d)), others can follow  peculiar
shapes (see Figure 7 (e)).}

 \vspace{0.3cm}

\hspace{0cm}\scalebox{0.60}{\includegraphics{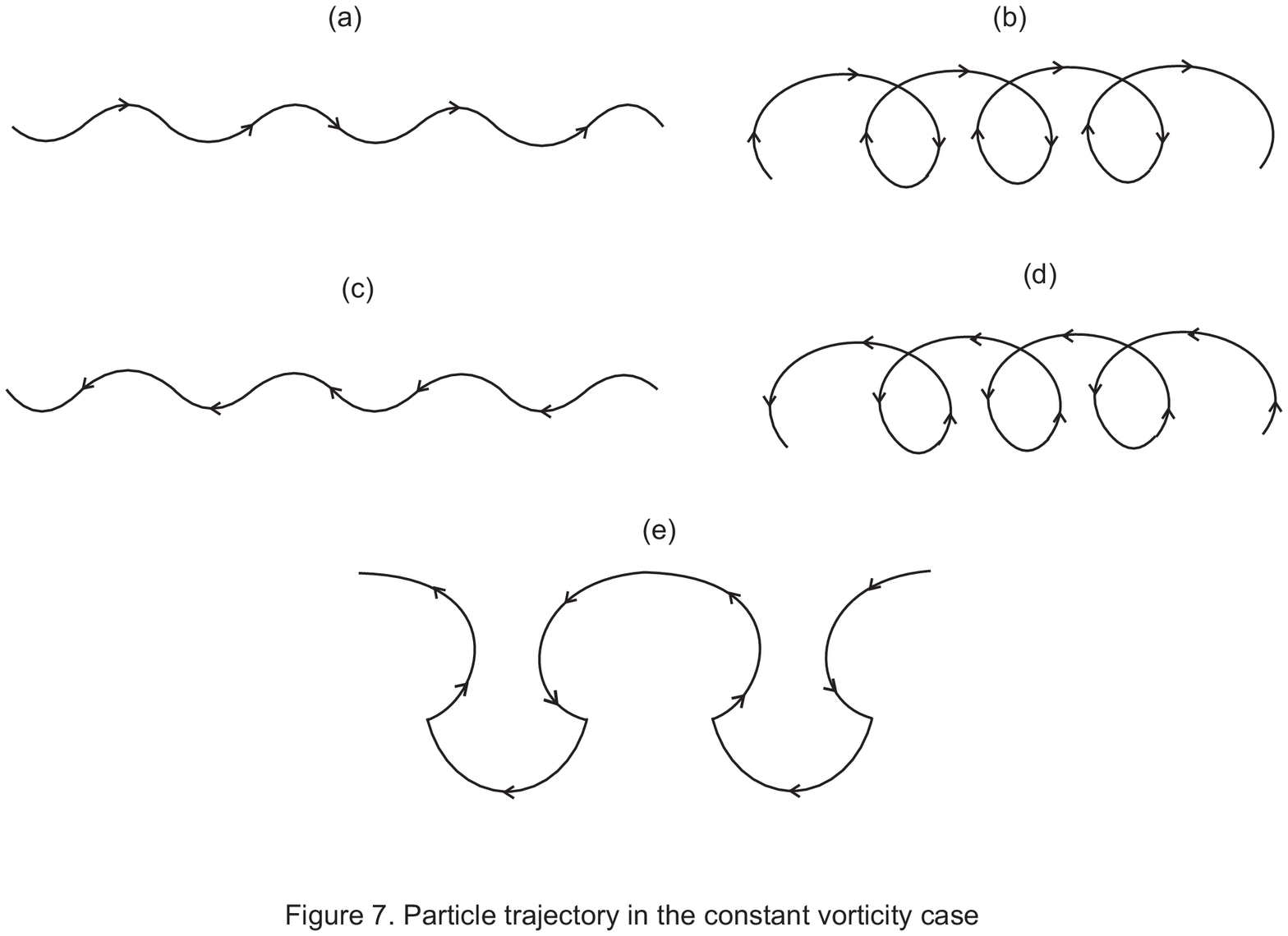}}
\\

\section*{Acknowledgments}
I would like to thank Prof. A. Constantin for helpful comments and
suggestions.

\medskip

\medskip

\end{document}